\documentclass[final,times,twocolumn]{elsarticle}
\usepackage{amssymb}
\usepackage{hyperref}
\usepackage{amsmath}
\usepackage{url}
\usepackage{marginnote}
\usepackage{xcolor}
\usepackage{soul}
\biboptions{sort&compress}
\usepackage{verbatim}
\usepackage{setspace}
\usepackage{enumitem}
\usepackage{graphicx}

\journal{Physics Letters B}

\newcommand{\df}{\mathrm{d}}

\newcommand{\nn}{\nonumber}

\begin{document}
\begin{frontmatter}

\title{\textsc{FastEEC}: Fast Evaluation of $N$-point Energy Correlators}

\author[Nikhef]{Ankita Budhraja}

\author[Nikhef,UvA]{Wouter J.~Waalewijn}

\affiliation[Nikhef]{organization={Nikhef, Theory Group},
              addressline={Science Park 105}, 
              postcode={1098 XG}, 
              city={Amsterdam} , 
              country={The Netherlands}}
            
\affiliation[UvA]{organization={Institute of Physics and Delta Institute for Theoretical Physics, University of Amsterdam},
            addressline={Science Park 904}, 
            postcode={1098 XH}, 
            city={Amsterdam},
            country={The Netherlands}}

\begin{abstract}
Energy correlators characterize the asymptotic energy flow in scattering events produced at colliders, from which the microscopic physics of the scattering can be deduced. This view of collisions is akin to analyses of the Cosmic Microwave Background, and a range of promising phenomenological applications of energy correlators have been identified, including the study of hadronization, the deadcone effect, measuring $\alpha_s$ and the top quark mass. 
While $N$-point energy correlators are interesting to study for larger values of $N$, their evaluation is computationally intensive, scaling like $M^N/N!$, where $M$ is the number of particles. In this Letter, we develop a fast, approximate method for their evaluation, exploiting that correlations at a given angular scale are insensitive to effects at other (widely-separated) scales. This implies that the energy correlator can be computed on (sub)jets, effectively reducing $M$. Furthermore, we utilize a dynamical (sub)jet radius that allows us to obtain reliable results without restricting the angular scales being probed. For concreteness, we focus on the projected energy correlator which projects onto the largest separation between the $N$ directions. E.g.~for $N=7$ we find a speed up of up to four orders of magnitude, depending on the desired accuracy. We also consider the possibility of raising the energy to a power higher than one in the energy correlator, which has been proposed to reduce soft sensitivity. These higher-power correlators are not collinear safe, but as a byproduct our approach suggests a natural method to regularize them, such that they can be described using perturbation theory. This Letter is accompanied by a public code that implements our method.
\end{abstract}


\end{frontmatter}

\section{Introduction}
\label{sec:introduction}

Energy correlators were proposed a long time ago in $e^+e^-$ collisions~\cite{Basham:1979gh,Basham:1978zq,Basham:1978bw,Basham:1977iq}. They received a lot of attention in recent years due to their extension to jets~\cite{Dixon:2019uzg,Chen:2020vvp,Lee:2022ige}. 
From a physics perspective, energy correlators naturally separate effects at different scales, providing a view of e.g.~the hadronization transition~\cite{Komiske:2022enw}, the dead-cone effect~\cite{Craft:2022kdo} and medium effects in heavy-ion collisions~\cite{Andres:2022ovj,Andres:2023xwr,Barata:2023zqg,Andres:2023ymw,Yang:2023dwc,Barata:2023bhh}. A range of other applications have been identified, such as the determination of the top quark mass~\cite{Holguin:2022epo,Holguin:2023bjf}, where the nature of e.g.~hadronization effects are different and (hopefully) better under control than for traditional observables. Indeed, energy correlators currently yield the most precise measurement at $\alpha_s$ from jet substructure~\cite{CMS:2024mlf}. 
For other recent phenomenological applications, see Refs.~\cite{Chen:2019bpb,Chen:2020adz,Chen:2021gdk,Chen:2022swd,Liu:2022wop,Liu:2023aqb,Cao:2023rga,Devereaux:2023vjz,Lee:2023npz,Lee:2024esz}.

Energy correlators were first studied on jets at hadron colliders using Open Data~\cite{Komiske:2022enw}, and have been measured at STAR~\cite{Tamis:2023guc}, ALICE~\cite{Mazzilli:2024ots} and CMS~\cite{CMS:2024mlf}.  
From a theoretical perspective, these energy correlators are described by a collinear factorization formula, and can also be described in terms of the expectation value of energy flow operators~\cite{Sveshnikov:1995vi, Tkachov:1995kk, Korchemsky:1999kt,  Hofman:2008ar, Belitsky:2013xxa, Belitsky:2013bja, Kravchuk:2018htv}. The latter has provided the opportunity to utilize  techniques developed in the context of conformal field theories to uncover the energy flow patterns in jets, resulting in numerous theoretical developments in the small angle limit~\cite{Dixon:2019uzg, Chen:2020vvp, Chen:2019bpb, Chen:2020adz, Chen:2021gdk, Schindler:2023cww, Gao:2023ivm, Chen:2023zlx, Chicherin:2024ifn, Chen:2024nyc, Chang:2022ryc, vonKuk:2024uxe,Kang:2023big,Lee:2023tkr}. Furthermore, the precision measurement of energy correlators is made possible due to the excellent performance of detectors at the Large Hadron Collider, where the tracking system plays a crucial role in accessing correlations at small angular scales. The calculation of energy correlators on tracks was developed in Refs.~\cite{Chen:2020vvp,Li:2021zcf,Chen:2022muj,Chen:2022pdu,Jaarsma:2023ell}, using the track function formalism~\cite{Chang:2013rca,Chang:2013iba}.

In this Letter, we focus on the projected $N$-point energy correlator (P$^{N}$EC), for which an operational definition is given by
\begin{align} \label{eq:EEC_def}
  \frac{\df \sigma_{N}}{\df R_L} &= \int \df \sigma \prod_{i_1, i_2, \dots, i_N}  z_{i_1}^\kappa z_{i_2}^\kappa \dots z_{i_N}^\kappa 
  \nn \\
  &\quad \times 
  \delta\bigl(R_L - \max\{\Delta R_{i_1,i_2},\Delta R_{i_1,i_3}, \dots \Delta R_{i_{N-1},i_N} \}\bigr)
\,.\end{align}
Here $\df \sigma_N$ is the differential cross section to produce some final state, $z_i = p_{T,i}/p_{T,{\rm jet}}$ is the momentum fraction of particle $i$ and 
\begin{equation}
\Delta R_{ij} = \sqrt{(\Delta y_{ij})^2 + (\Delta \phi_{ij})^2}
\end{equation}
is the distance between particles $i$ and $j$ in the rapidity-azimuth $(y, \phi)$ space. 
The delta function in Eq.~\eqref{eq:EEC_def} picks out the largest distance between the particles $i_1, \dots i_N$. 
At the leading logarithmic accuracy, it has been shown that for the cumulative distribution defined as
    \begin{equation}
    \Sigma_N = \int_0^{R_L} {\rm d}R_L^\prime \frac{{\rm d}\sigma_N}{{\rm d}R_L^\prime} \, ,
    \end{equation}
    this projection to the largest distance exhibits a simple (approximate) scaling behavior $\Sigma_N \sim R_L^{\gamma(\alpha_s,N)}$. 
    The exponent of this scaling is related to the DGLAP anomalous dimensions, specifically the $N$-th moments of the QCD splitting functions.
The default choice for the power (weight) $\kappa$ is 1, but we will also discuss other choices.
In principle, one can also constrain more than just the largest distance. In that case the energy correlator would depend on up to $\tfrac12 N(N-1)$ distances, but at this point it is not clear which of these are most relevant.

\begin{figure}
    \centering
	\includegraphics[width=0.48\textwidth]{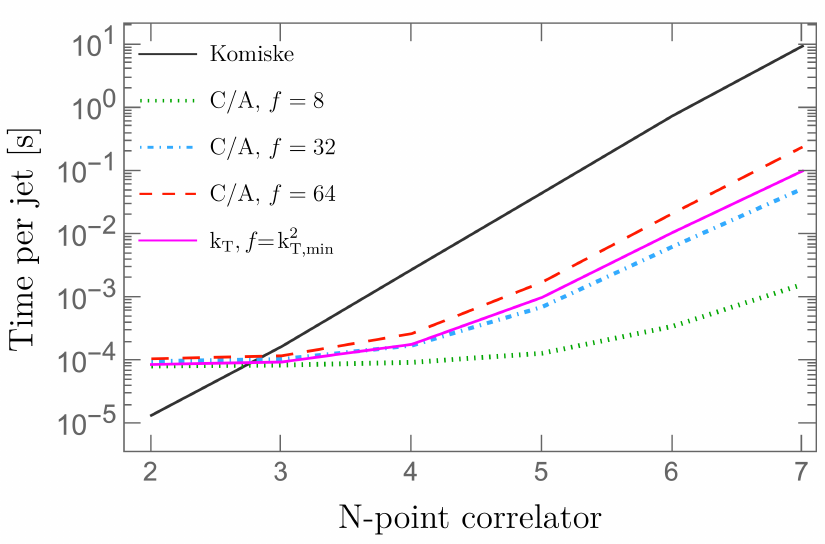}	
 \caption{Average time per event using Ref.~\cite{EEC_github} and our fast method, employing C/A or $k_T$ to recluster  the jet and the indicated resolution factor $f$. While all show exponential scaling with $N$, their slopes are different leading to 
 orders of magnitude improvement at large $N$, depending on the desired accuracy.}
 \label{fig:time}
\end{figure}

As is clear from Eq.~\eqref{eq:EEC_def}, the calculation of the $N$-point correlator for a final state with $M$ particles scales like $M^N/N!$ with $M \gg N$, where the factorial arises because the ordering of $i_1$, $i_2$, \dots $i_N$ is irrelevant. For large values of $N$ this becomes prohibitive, and indeed studies so far have been restricted to $N\leq6$, with $N=6$ already requiring substantial computational overhead~\cite{Komiske:2022enw, Jaarsma:2023ell}. We address this problem here, finding that a substantial speed up is possible, depending on the desired level of accuracy. To pique your interest, we show in Fig.~\ref{fig:time} the average time per event using our approach as function of $N$, compared to the current \verb+EnergyEnergyCorrelators+ package~\cite{EEC_github}. This clearly shows a speed up of multiple orders of magnitude, depending on the desired precision (which is controlled by the resolution parameter $f$ discussed later). The basic idea that we use is that correlations at a given scale are insensitive to details at much smaller scales, that don't need to be resolved, thus reducing $M$. Furthermore, radiation separated by larger scales can be treated as independent, e.g, if the $M$ particles consist of two well-separated clusters with $m$ and $M-m$ particles, this replaces $M^N$ by $m^N + (M-m)^N$ for correlations at distances smaller than the separation between the two clusters.

In Eq.~\eqref{eq:EEC_def} also $\kappa>1$ has been considered, as this further suppresses the contribution from soft radiation (see e.g.~\cite{Holguin:2022epo,Barata:2023bhh}). In this case, the energy correlator is not collinear safe, making it much more sensitive to hadronization physics. This distinction is not so relevant for us, as we simply focus on speeding up the calculation of an energy correlator for a given final state. Interestingly, our fast method for calculating energy correlators suggests a natural way to restore collinear safety, and thereby their perturbative calculability, without limiting the range of $R_L$.

The structure of this Letter is as follows, in Sec.~\ref{sec:method} we discuss the basic principles behind our method, as well as the (dis)advantages of various choices (reclustering and resolution parameter) in its implementation. Sec.~\ref{sec:implementation} discusses the basic usage of our public code~\cite{FASTEEC}, that accompanies this Letter. We show our numerical results in Sec.~\ref{sec:results} for $N=2$ through 7, 
and our conclusions and outlook are in Sec.~\ref{sec:conclusions}.

\section{Fast evaluation of energy correlators}
\label{sec:method}

The underlying idea that we utilize is as follows: for correlations at a given separation $\Delta R$ in $(y,\phi)$-space, radiation that is much closer together can be clustered and treated as one. 
The simplest way to achieve this would be to use a jet algorithm, effectively reducing the number of particles by clustering them into subjets. However, this only allows one to calculate the energy correlator at angles (much) larger than the subjet radius, $R_L > r$. To access correlations at small angles would require reducing the subjet radius, thereby losing the computational speed up. This can be remedied by noting that  radiation separated by distances larger than $\Delta R$ can be treated independently, which our approach takes advantage of. 

\begin{figure}
       \centering
       \includegraphics[width=0.45\linewidth]{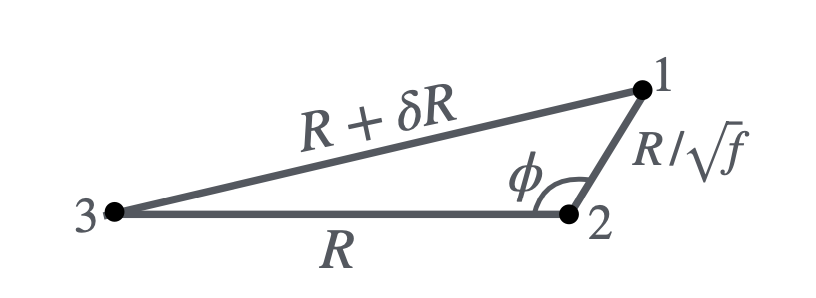}
       \caption{A pictorial representation of two closely spaced particles with angular
       orientation $\phi$.}
        \label{fig:triangle}
    \end{figure}

\begin{figure}
    \centering
	\includegraphics[width=0.48\textwidth]{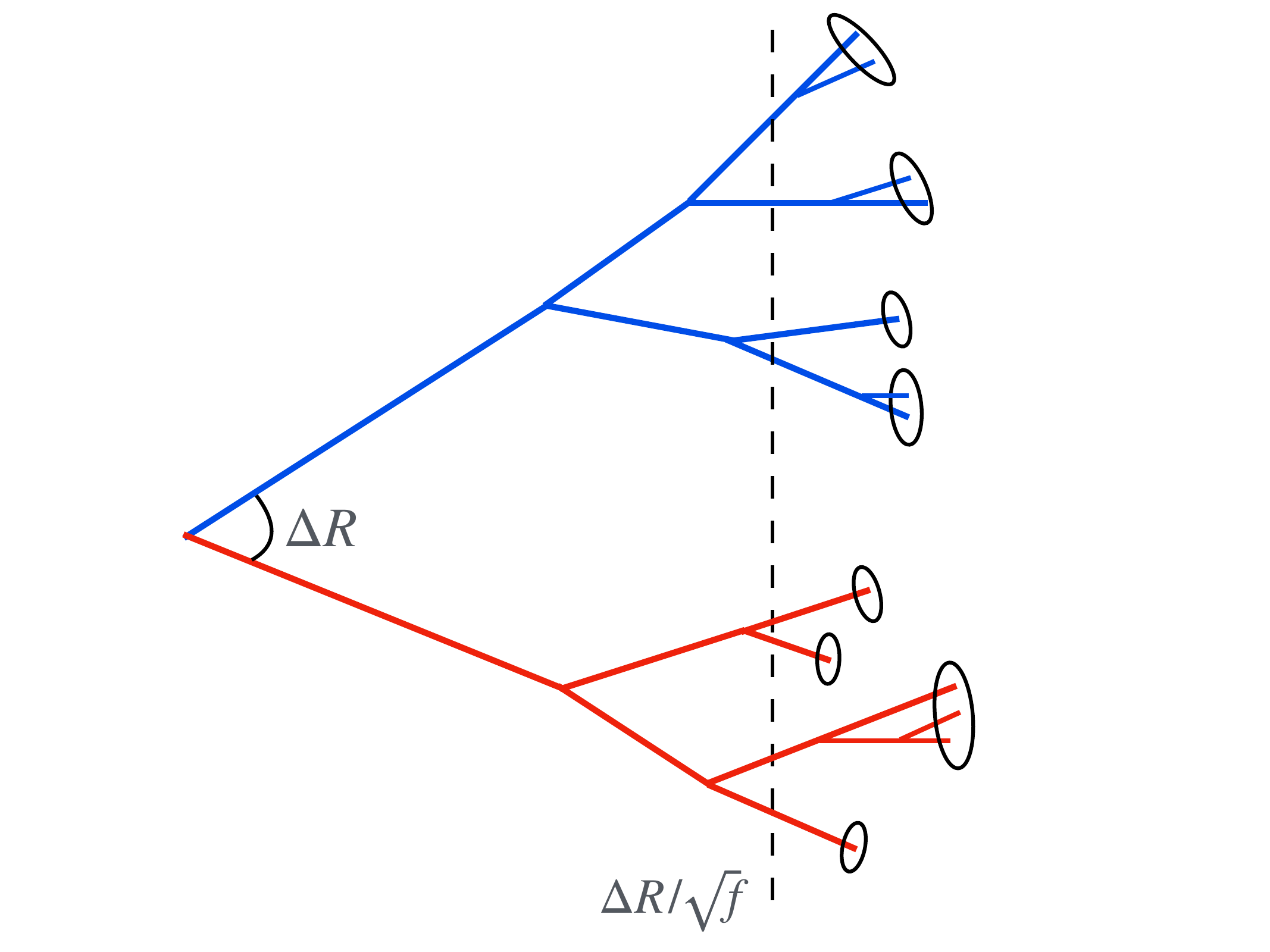}	
 \caption{Pictorial representation of the fast algorithm. Particles with angular separation smaller than $\Delta R/\sqrt{f}$, where $f$ is the resolution factor, get clustered into subjets. This approximation is justified for contributions involving both branches (blue \emph{and} red), while contributions involving a single brach are obtained recursively.}
 \label{fig:method}
\end{figure}

We use a dynamic resolution scale, such that for a given scale $\Delta R$, a subjet radius $r = \Delta R/\sqrt{f}$ is employed with $f>1$ the resolution parameter. Rather than sampling over all possible values of $\Delta R$, it is convenient to use the $\Delta R$ separations present in the particles in the jet. Concretely we achieve this by  reclustering the jet using Cambridge/Aachen (C/A)~\cite{Dokshitzer:1997in}, which results in an angular-ordered clustering tree. We then recursively traverse the tree, using the angle between  two branches as $\Delta R$ and resolving the two branches using a subjet radius $r = \Delta R/\sqrt{f}$, see Fig.~\ref{fig:method}. We can then calculate the contributions to the energy correlator involving both branches, i.e.~some blue and some red particles. Here the separation $\Delta R$ dominates the distance, justifying the the use of subjets as a good approximation. To also obtain the contribution involving particles from only one branch (only blue or only red) we repeat this approach recursively on each of the branches.  It is instructive to note that this approach can be thought of as a Taylor expansion of the exact result as follows: Consider two particles ($1$ and $2$) located at distances $R$ and $R+\delta R$ from a third particle $3$ (see Fig.~\ref{fig:triangle}) in the (\(y-\phi\)) space with an angular orientation specified by $\phi$ such that $\delta R = -\frac{R}{\sqrt{f}}\cos\phi + \frac{R}{2f}\sin^2\phi + {\cal O}(f^{-3/2})$.  
These particles are then combined to form a subjet instead of being treated separately. Concretely, this implies that for  correlations at a given angular scale $R > \delta R$, the result can be approximated as
    \begin{equation}
    w(R+\delta R) \approx 
         w(R) 
        + w^\prime(R)\, \delta R \, 
        + {\cal O}[(\delta R)^2]\, ,
    \end{equation}
    where $w(R+\delta R)$ is the energy correlator corresponding to the exact computation and $w(R)$ represents the approximate result due to the particles at $R$ and  $R+\delta R$, treated as one. The leading correction in $\delta R$ that goes as $-(R/\sqrt{f}) \, \cos\phi$ averages out to zero, so the extent of the relative error (due to the approximation) is governed by $1/f$. Larger values of $f$ will increase the accuracy but impact the computational time and we will show that $f=32$ already yields results at about $1-2\%$ precision with an evaluation time up-to two orders faster than the exact computation, for $N=7$.
In our implementation we use the \textsc{FastJet} package~\cite{Cacciari:2011ma,Cacciari:2005hq} for jet reclustering and the subsequent resolution of its substructure in terms of subjets. 

We now describe our method more systematically, illustrated in Fig.~\ref{fig:method}:
\begin{enumerate}
    \item Recluster the final state particles into a jet using the C/A (or $k_T$) algorithm~\cite{Dokshitzer:1997in,Catani:1993hr}. We choose a sufficiently large jet radius $R=1.5$  for this clustering step so that all the particles are inside one jet.
   \item Consider the first split, for which the parent branches (shown in blue and  red) of the reclustering tree at an angular distance $\Delta R$  are identified. Decluster each branch into subjets (shown as ellipses) with radius parameter $r = \Delta R/\sqrt{f}$.
    \item Calculate the contribution to the energy correlator using these subjets, \emph{restricting} to contributions involving at least one subjet from each branch.
  \item Return to step 2 for \emph{each} of the two branches to calculate the contributions involving particles on one branch. Recurse until branches contain a single particle.
\end{enumerate}

We have explored various combinations of jet algorithms to recluster the jet and resolution factors $f$. Our default is to recluster using Cambridge/Aachen with a fixed value for  $f$. As we will see in Sec.~\ref{sec:results}, the accuracy of our predictions (compared to the full calculation) is not constant as function of the angular scale. One can make the accuracy (more) constant by choosing a value of $f$ that depends on $\Delta R$. We have explored this and found that it can speed things up by a factor of about 2, but didn't include this functionality as a standard option in our code.

 \begin{figure}[t]
        \centering
        \includegraphics[width=0.45\textwidth]{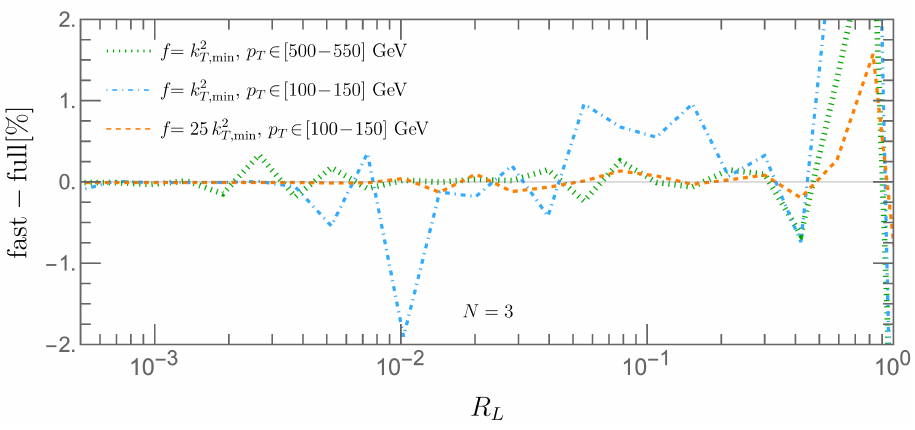}
        \includegraphics[width=0.45\textwidth]{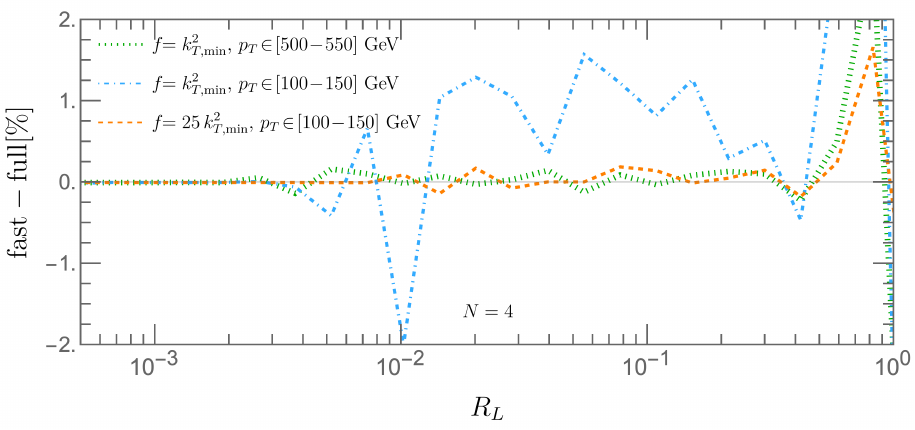}
        \caption{\footnotesize{ To determine the scaling of $f$ with the jet $p_T$ for our $k_T$-based approach, we try two different choices: $f = k_{T,{\rm min}}^2/(1\,{\rm GeV})^2$ and $f=k_{T,{\rm min}}^2 \times (500/p_{T})^2$. These agree for $p_T = 500 GeV$ but differ for other values of $p_T$, and we show the difference with the full result for $N=3$ (upper) and $N=4$ (lower). Clearly the second choice is the better one, with stable performance for different $p_T$ bins.}}
        \label{fig:kt}
    \end{figure}

Reclustering with the $k_T$ algorithm performs worse when using a constant resolution $f$: to get a similar level of accuracy requires a much larger $f$ and thus much more computing time. 
Interestingly, in this case, we find that using $f=f^\prime k_{T,{\rm min}}^2$ with $f^{\prime}$ some dimensionful scale (in units of GeV$^{-2}$), yields rather good results, with about a per-mille level precision for all $N$ values, within a very reasonable amount of time. When using $f= f^\prime k_{T,{\rm min}}^2$, the algorithm dynamically resolves the subsequent substructure of the splittings depending on the transverse momenta of the partons, thereby resolving more for splittings with large $k_{T,{\rm min}}$. 
     Further, in order to determine the dependence on the jet $p_T$ of  the dimensionful $f^\prime$, we perform Monte Carlo simulations and generate jets in two different $p_T$ ranges i.e. $p_T \in [500-550]\, {\rm GeV}$ and $p_T \in [100-150]\, {\rm GeV}$ using the anti-$k_{T} $ algorithm~\cite{Cacciari:2008gp} with a radius of $R=1.0$. Using two different choices: $f=k_{T,{\rm min}}^2/(1 {\rm GeV})^2$ and $f=k_{T,{\rm min}}^2\times(500/p_{T, {\rm jet}})^2$ that coincide for $p_{T,{\rm jet}}=500\, {\rm GeV}$, we study the relative error for the low $p_T$ bin and $N=3,4$. By comparing the relative error for two different jet $p_T$ ranges, from Fig.~\ref{fig:kt}, we find that using $f^\prime \sim (500/p_{T,{\rm jet}})^2$ has little sensitivity across the two different $p_T$ ranges.

Note that anti-$k_{T}$~\cite{Cacciari:2008gp} does not improve the implementation, because anti-$k_{T}$ first clusters the energetic radiation in the jet, adding all soft radiation close to the jet boundary at the end. Consequently, declustering the branches of the first split yields many subjets, leading to very large computation times.

Additionally, our method can also be applied to compute the projected correlator where transverse momenta in Eq.~\eqref{eq:EEC_def} are weighted with power $\kappa > 1$. Although, the higher power of transverse momentum weighting is collinear unsafe, it has been e.g.~proposed to help mitigate the overwhelming underlying event activity in the complex environment of heavy ion collisions~\cite{Barata:2023bhh}, or improve the resolution with which the top quark mass can be extracted~\cite{Holguin:2022epo}. In this case, the transverse momentum of the subjet used to calculate the energy correlator should be taken as $(\sum_{i \in \text{subjet}} p_{T,i}^\kappa)^{1/\kappa}$ and not $\sum_{i \in \text{subjet}} p_{T,i}$, in order to agree with the full calculation of the energy correlator.

\section{Implementation}
\label{sec:implementation}

Our code implements the fast algorithm discussed in the previous section utilizing the \textsc{FastJet} package for reclustering jets and resolving their substructure~\cite{Cacciari:2011ma,Cacciari:2005hq}. Four flavors of the code are made available, using respectively C/A and $k_T$ reclustering, and taking the transverse momenta of the particles to the power $\kappa = 1$ and $\kappa \neq 1$.

We start by discussing the code with C/A for $\kappa=1$. When the code is executed from the command line, the user is required to pass: the inputfile from which events should be read (\verb+input_file+), the total number of jets to be analyzed\footnote{Note that we assume a jet, rather than an event, as input.}  (\verb+events+), $N$ that specifies which point correlator to compute, the resolution factor $f$, the minimum bin value (\verb+minbin+) of the histogram, the total number of bins (\verb+nbins+) of the histogram and the output filename (\verb+output_file+). In short, the command line syntax is: 
\begin{align*}
\texttt{
./eec\_fast input\_file events $N$ $f$ minbin} \\
\texttt{nbins output\_file}
\end{align*}
The minimum bin value \verb+minbin+ should be entered as a $\log_{\rm 10}(R_L)$ number. We have fixed the maximum bin value to be 0 in these units, corresponding to $R_L=1$. The output of the generated histograms is normalized because of momentum conservation, and the lowest (highest) bin include the underflow (overflow). We currently support up to $N=8$ for higher point projected correlators, but this can easily be extended to higher values as well, if needed. 

In our default C/A version, we support constant jet resolution values of $f > 1$, where larger values of $f$ yield more accurate results but require more time. For the $k_T$ reclustering case, the program is called \texttt{eec\_fast\_kt}. In this case, $f = f' k_{T,{\rm min}}^2$, where $f'>0$ is now the command line parameter.

The code in which the transverse momenta of the particles are taken to a power $\kappa \neq 1$, is called \texttt{eec\_fast\_weight} and \texttt{eec\_fast\_kt\_weight}. It takes $\kappa>0$ as an additional command line parameter. Because this energy correlator is not collinear safe, we need to add the transverse momenta of the constituents of a subjet as $\sum_{i \in \text{subjet}} p_T^{\kappa}$, but we can  treat the constituents as moving in the same direction, maintaining the desired speed up.

The first line of the output file consists of \verb+events+, \verb+nbins+, \verb+minbin+, \verb+maxbin+, where 
\verb+maxbin+ $= \log_{10}(1) = 0$. The next line contains the histogram values, and the number 
of entries equals \verb+nbins+.  We have also included a small \textsc{Mathematica} notebook along with our public code, that illustrates how output files can be read~\cite{FASTEEC}. We illustrate this for the 4-point energy correlator, and include the necessary output files to reproduce the corresponding panel of Fig.~\ref{fig:ENC}. Below we discuss in detail, the performance of our fast method and the relative errors associated to different approximations outlined above.

\section{Results and performance}
\label{sec:results}

The results presented here are based on the publicly available "MIT Open Data" (MOD)~\cite{Komiske:2019jim,MITOpenData} which utilizes the reprocessed data on jets from the CMS 2011A Open Data~\cite{CERNOpenDataPortal,CMSOpenData}. This dataset consists of jets with a transverse momenta $p_{T} \in [500,550]$ {\text GeV} and pseudo-rapidity $\vert \eta \vert < 1.9$. For our analysis, we have converted this dataset from MOD format to a text file format, that on each line lists: jet number, transverse momentum $p_{T}$, rapidity $y$ and azimuthal angle $\phi$ of a particle in that jet. This input file is also made available along with our public code and contains a total 100\,000 jet events in the specified $p_{T}$ range. Note that we construct the correlator on all the final state particles in the jet and not only on charged hadrons. No detector effects or pileup subtraction is performed, as this is not the focus of our study.

\begin{figure*}
	\centering 
	\includegraphics[width=0.48\textwidth]{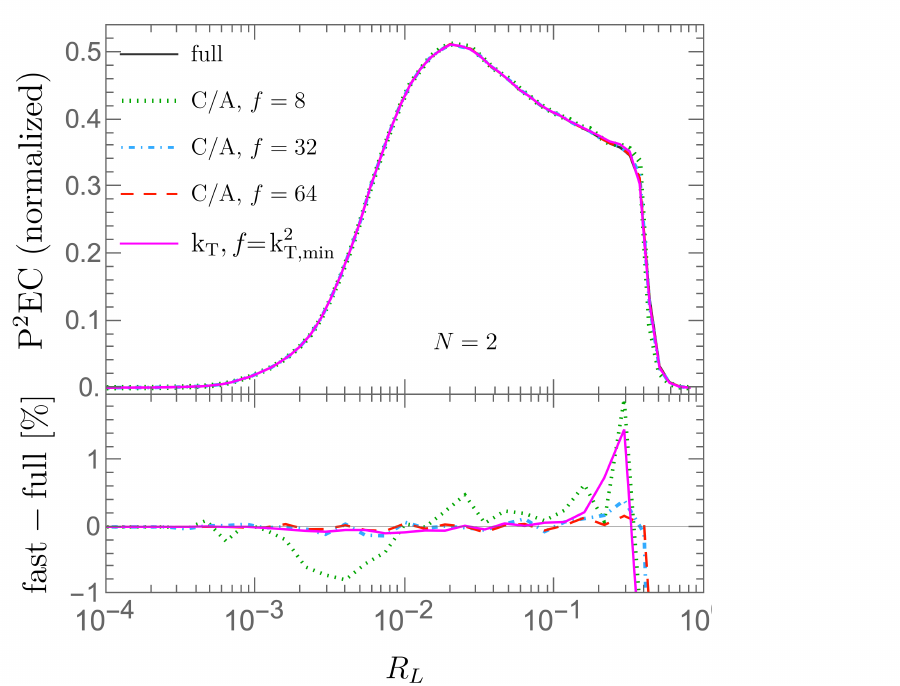}	
	\includegraphics[width=0.48\textwidth]{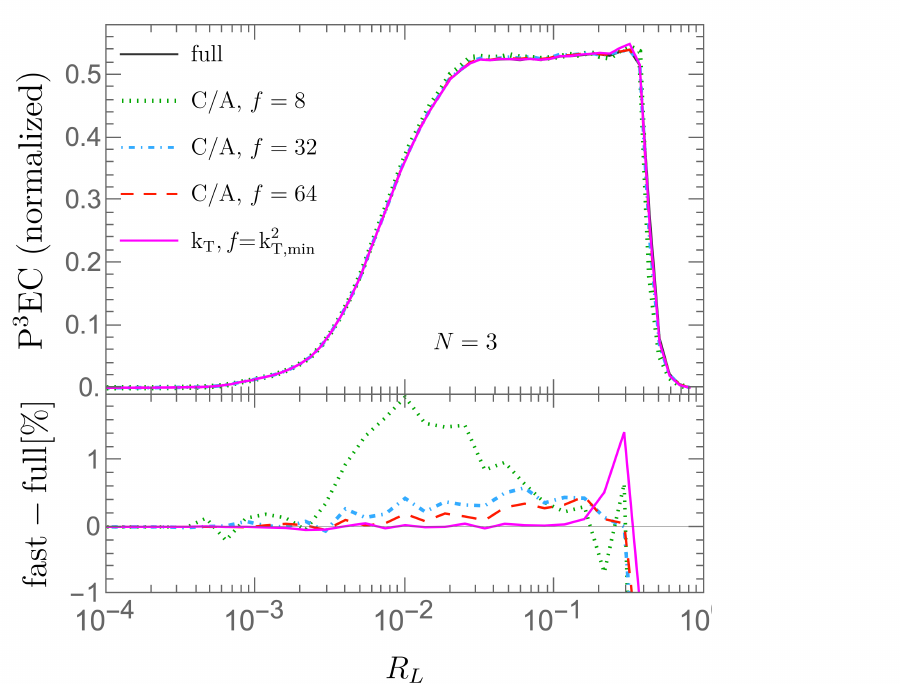}	\\
	\includegraphics[width=0.48\textwidth]{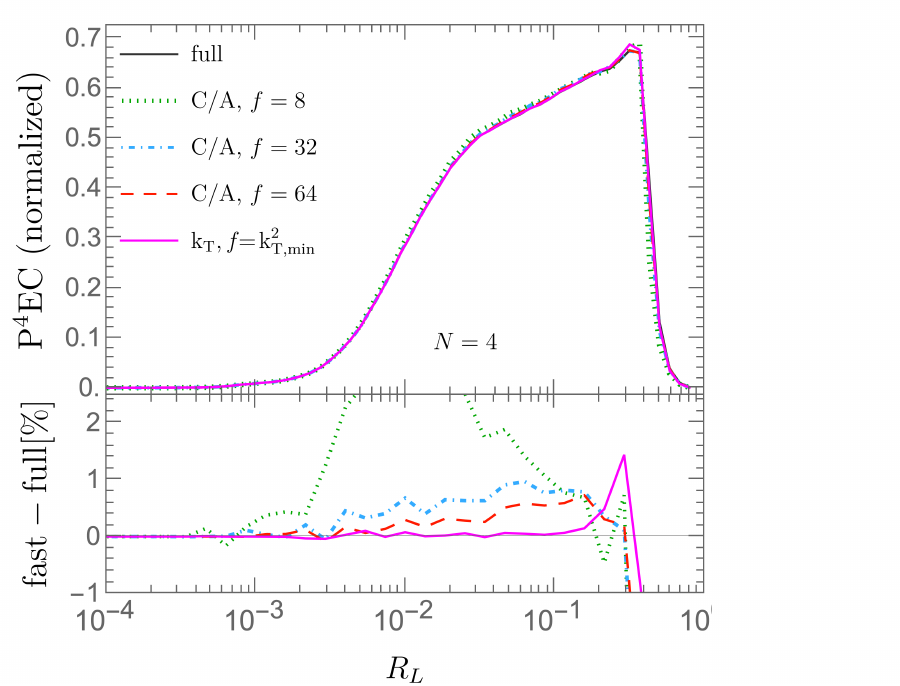}	
	\includegraphics[width=0.48\textwidth]{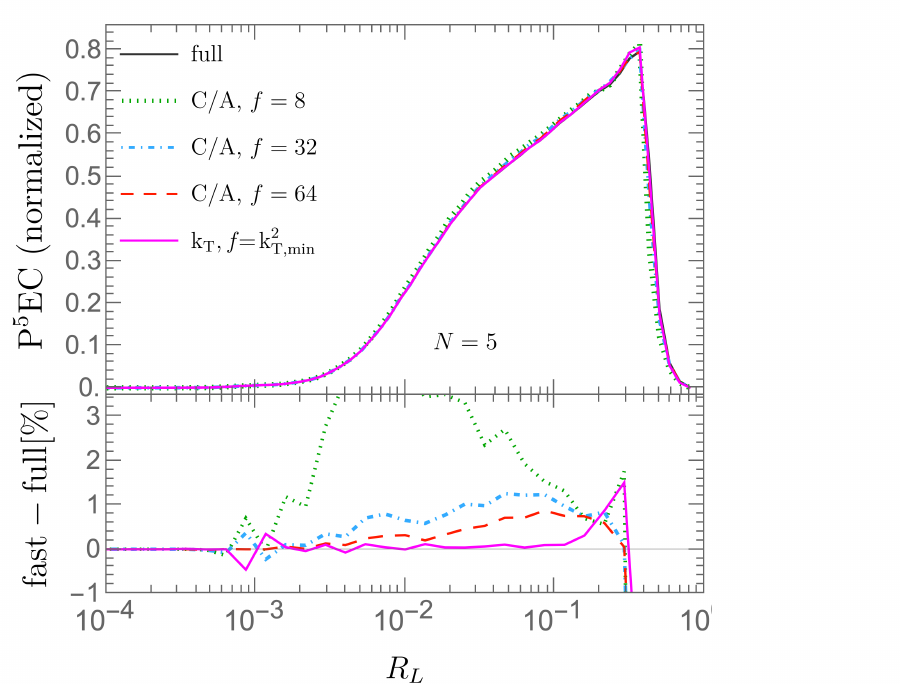}	\\
	\includegraphics[width=0.48\textwidth]{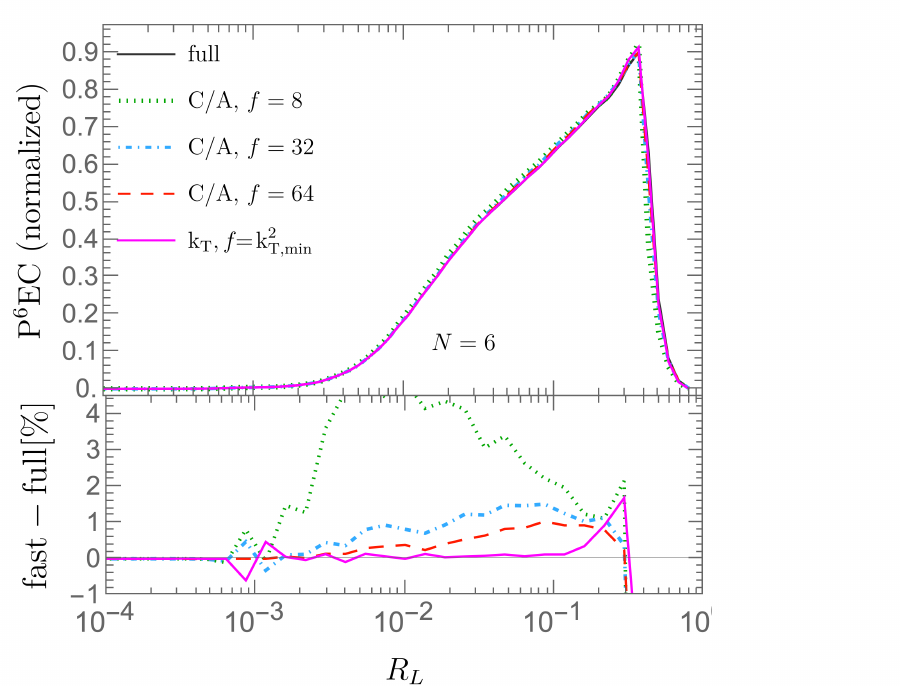}	
	\includegraphics[width=0.48\textwidth]{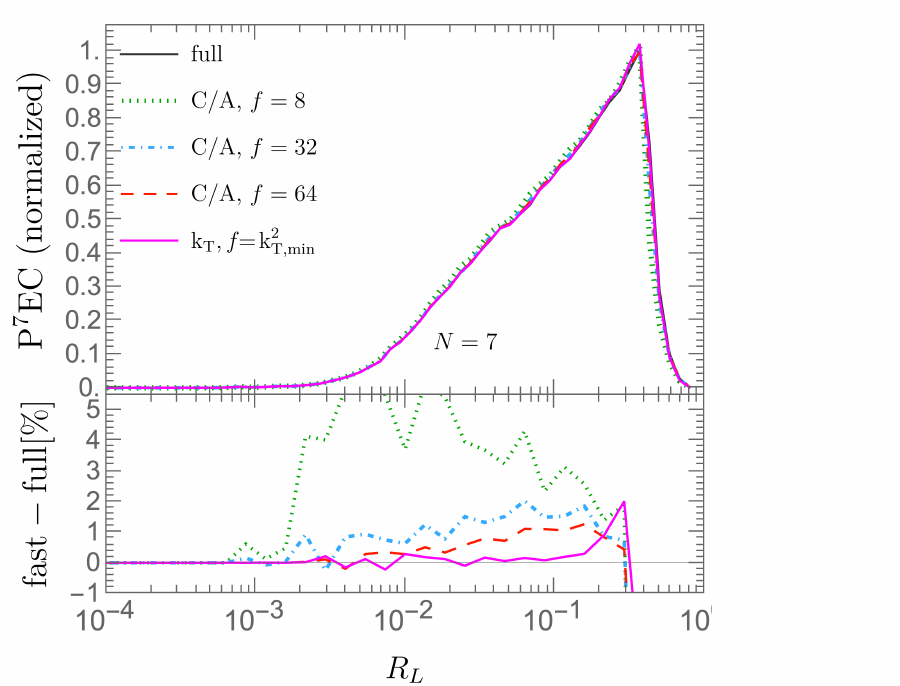}	\\
 \caption{Projected $N$-point energy correlator for $N=2$ to 7, calculated using Ref.~\cite{EEC_github} and our fast method, employing C/A or $k_T$ to recluster the jet and the indicated resolution factor $f$. In the lower panel the relative error of these methods are shown, where we halved the number of bins to smoothen out the statistical fluctuations a bit. While the error from using $f=8$ increases from 2\% to 5\% as $N$ increases from 2 to 7, which can be counteracted by increasing $f$, the $k_T$ reclustering with $f=k_{T,{\rm min}}^2$ performs well across the board.}
	\label{fig:ENC}%
\end{figure*}

\begin{figure*}
	\centering 
	\includegraphics[width=0.48\textwidth]{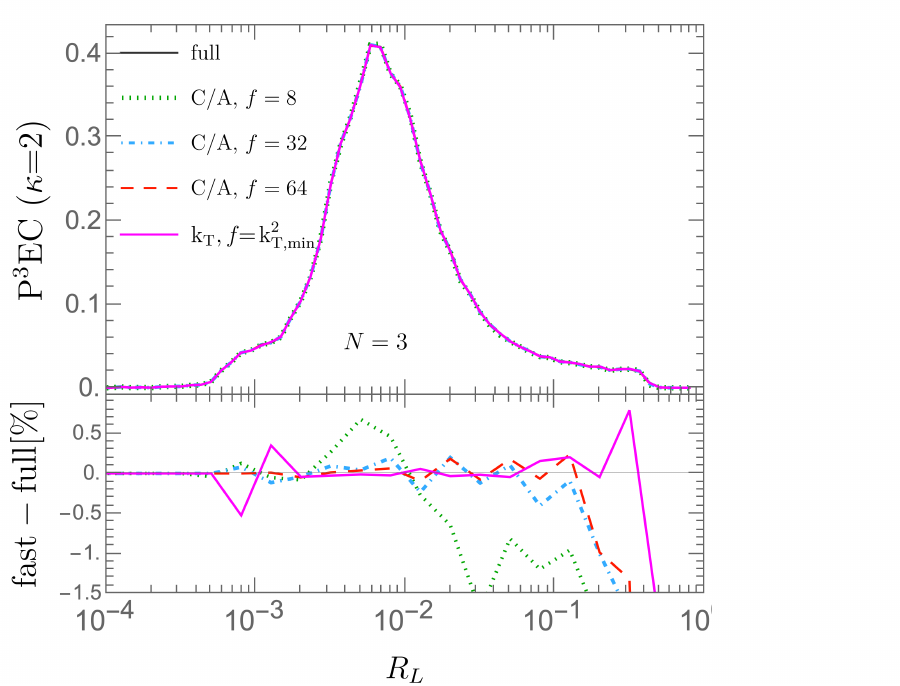}
	\includegraphics[width=0.48\textwidth]{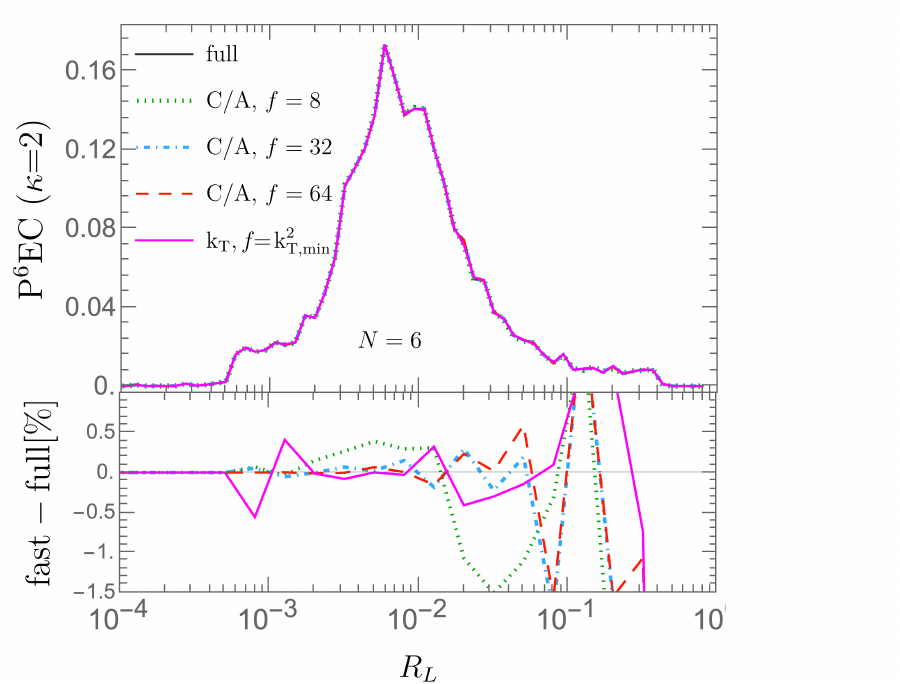}
 \caption{Projected $N$-point energy correlator with $\kappa = 2$ for $N=3$ (left) and $N=6$ (right), calculated using Ref.~\cite{EEC_github} and our fast method, employing C/A or $k_T$ to recluster the jet and the indicated resolution factor $f$. In the lower panel the relative error of these methods are shown, where we have reduced the number of bins by a factor three to smoothen out statistical fluctuations. We see that in this case the error from using $f=8$ is about 1.5\% in both the cases which improves down to less than 0.5\% with $f=32$. We also find that the $k_T$ clustering with $f=k_{T,{\rm min}}^2$ gives a maximum error of about 0.5\% as well, though this is much smaller for the $N=3$ case.}
	\label{fig:EkC}%
\end{figure*}

The results for our projected $N$-point correlators were obtained with a single core on a M2 chip MacBook Air both for the full calculation, for which we use the publicly available \verb+EnergyEnergyCorrelators+ package~\cite{EEC_github}, and using our fast computation method. Fig.~\ref{fig:time} shows the average computation time per event (in seconds) that is used for the full calculation, and each of the different choices of reclustering algorithm and resolution factor $f$ described earlier, up to the $N=7$ point correlator. 
We find that our method provides a substantial gain in the computation time of higher point correlators, with $N=5$ already achieving a gain by a factor of about 20, even using the slowest setting C/A $f=64$ among the choices we considered. For correlators with $N>5$, this gain is even more substantial, reaching up to a factor of about 40 for N=7 for C/A with $f=64$. The fastest option we considered, C/A with $f=8$ has a speed up of a factor 5000 for $N=7$, corresponding to an evaluation time of only about $10^{-3} s$ per jet.  Additionally, we observe that for the case of $k_T$ clustering with $f=k_{T, min}^2$, the algorithm provides an improvement by a factor 2 for $N=7$ when compared to C/A with $f=64$ clustering.
We note that even though for $N=2$ our method does not perform as efficiently as Ref.~\cite{EEC_github} (see Fig.~\ref{fig:time}), the use of physically-motivated approximations in our approach allows for the substantial gains we observe for $N > 3$.

The distribution for the $N$ point correlator obtained by using different clustering schemes and choices of the resolution factor is shown in Fig.~\ref{fig:ENC}, along with the relative error of these methods when compared to the full computation. We show results for  $N= 2 $ to $7$ point projected correlators, and the distributions are normalized such that the area under the curve is unity. For all the distribution plots shown in this Letter, we create histograms between $10^{-5} < R_L < 1$ using a total of 75 logarithmic bins. Note that for the error plot, we halved the number of bins to reduce statistical fluctuations in the plot. 

There is only one region in the plot where these approximations do not perform so well, corresponding to the edge of the plot. This is not surprising, as the fine details of radiation at the jet boundary will be important to accurately describe this, making it sensitive to the jet algorithm. However, this is not the region of interest since in most applications of energy correlators one is interested in (much) smaller scales. One concrete application is the determination of the power-law scaling in the perturbative region, which is directly related to the moments of the time-like splitting functions. Using the CMS Open Data, it was shown in Ref.~\cite{Komiske:2022enw} that the low-point ($N=2-6$) energy correlator distribution exhibits different power law scalings as a function of the angular separation between the final state particles. In particular, 
away from the jet boundary, there is a perturbative power law scaling governed by splitting functions and at even smaller angles there is another power-law that can be interpreted as a free-hadron gas~\cite{Komiske:2022enw}.

The approximations we make could, therefore, impact the power-law behaviour (in the perturbative region) extracted from LHC data. However, as is clear from Fig.~\ref{fig:ENC}, our method can still be used to reliably extract this power-law scaling at the few percent-level or less with a reasonable computation time. This implies that our method provides access to this scaling for much larger values of $N$ than currently realized through the existing \verb+EnergyEnergyCorrelators+ package due to the computation time.

We find that in the region of interest,  the relative error when using C/A with $f=8$ grows from about 1\%  at $N=2$ to 5\% for $N=7$. This can be remedied by choosing a larger value of $f$, and for $f = 64$ the error is still below 1\% for $N=7$. Interestingly, we observe that using $k_T$ clustering with $f=k_{T,min}^2$ has a much smaller error of a per mille or less over most of the range, while its time is between C/A with $f=32$ and $f=64$. 

Next, we also study the behavior of correlators when higher powers of transverse momentum weights are used. Specifically, we present results for the case when $\kappa =2$, which is shown in Fig.~\ref{fig:EkC} for $N=3$ and $N=6$, along with the relative error. The distributions are again normalized such that the area under the curve is unity. Note that being collinear unsafe seems to lead to much larger statistical fluctuations (the jet sample is the same as in Fig.~\ref{fig:ENC}). Consequently, we reduce the number of bins in the relative plot by a factor of 3, to make the trends more visible.

\section{Conclusions and outlook}
\label{sec:conclusions}

In this Letter, we focused on the projected $N$-point energy correlator and proposed a method that provides a substantial speed up in the computation of higher point correlators. The underlying idea that we employ is that for correlations at a certain angular scale, radiation separated by much smaller distances can be treated as one, and radiation separated by larger distances can be treated as independent. We achieve this by reclustering with C/A or $k_T$, recursing over the tree, and using a dynamical subjet radius $r = \Delta R/\sqrt{f}$, with $\Delta R$ the separation between the two parents of the split under consideration. While C/A with fixed $f$ requires increasing the resolution to maintain accuracy for larger values of $N$, recustering with $k_T$ and using $f=k_{T,min}^2$ provides excellent accuracy for all higher-point correlators we studied. The gain in speed we obtain is one or more orders of magnitude,  depending on the desired accuracy. We also utilize our method for the case where higher powers of transverse momenta are taken.

There are several interesting avenues to extend the method proposed in the Letter. First, this method can be straightforwardly extended to the case of second-largest separation $R_S$ between the $N$-directions, since $R_S > R_L/(N-1)$ and is thus parametrically of the same size. Second, this method can also be taken as the \emph{definition} of an observable, whose leading-logarithmic (LL) calculation is the same as that of the projected energy correlator~\cite{Dixon:2019uzg}, since  angles are \emph{strongly} ordered in this limit.  Finally, for collinear unsafe observables, one can use subjets to regulate the collinear divergences. However, this implies that there is no sensitivity below the scale of the subjet radius. The method described here, provides another way of regulating these divergences that does not limit the scales that are probed.\footnote{If the resolution parameter $f$ is not constant, as in the $k_T$ example, there can be large sensitivity to hadronization effects throughout the distribution.}
An alternate way to achieve this is to use Lund-plane based clusterings, as proposed in Ref.~\cite{Barata:2023bhh}. Note that our current implementation for $\kappa \neq 1$ does not do this, because we want to reproduce the collinear-unsafe full calculation of the EEC.

\section*{Acknowledgements}
We thank S.~Alipour-fard, E.~Chasapis, M.~Jaarsma and I.~Moult for discussions. This publication is supported by EU Horizon 2020 research and innovation programme, STRONG-2020 project, under grant agreement No 824093.

\bibliographystyle{elsarticle-num}
\bibliography{main}

\end{document}